# Intrinsically ultralow thermal conductivity in all-inorganic superatomic bulk crystals

Mingzhang Yang,[1,2,‡] Yuxi Wang,[3,‡] Jun Deng,[4] Tianping Ying,[1] Qinghua Zhang,[1] Nianjie Liang,[3] Xiaobing Liu,[5] Bai Song,[3,a] Jian-gang Guo,[1, a] Xiaolong Chen[1,a]

[1]Beijing National Laboratory for Condensed Matter Physics, Institute of Physics, Chinese Academy of Sciences, Beijing 100190, China

[2]School of Materials Science and Opto-Electronic Technology, University of Chinese Academy of Sciences, Beijing, 101408, China

[3]Department of Energy and Resources Engineering, Peking University, Beijing 100871, China.

[4]Center for High Pressure Science and Technology Advanced Research, Beijing, China

[5]Key Laboratory of Quantum Materials under Extreme Conditions in Shandong Province, School of Physics and Physical Engineering, Qufu Normal University, Qufu 273165, China

‡These authors contribute equally.

a)Authors to whom correspondence should be addressed: songbai@pku.edu.cn; jgguo@iphy.ac.cn; xlchen@iphy.ac.cn

ABSTRACT

Superatomic compounds, composed of atomic clusters interwoven by weak chemical bonds exhibit large anharmonicity vibrations, are excellent candidates for ultralow thermal conductivity ($\kappa$) materials. However, growing bulk superatomic single crystals is challenging due to complex chemical composition and chemical bonds, and studies on their intrinsic thermal property are scarce. Here, we grew high-quality superatomic single crystals of $Re_6Se_8Te_7$ and $Re_6Te_{15}$, both of which are narrow band gap semiconductors that change into metals under external physical pressure. At room-temperature, the $\kappa$ are 0.32 W $m^{-1}$ $K^{-1}$ and 0.53 W $m^{-1}$ $K^{-1}$ in $Re_6Se_8Te_7$ and $Re_6Te_{15}$, respectively, ranking among the lowest value reported in all-inorganic bulk crystals. It is mainly attributed to the large Grüneisen parameter (1.93) and low average sound speed (< 1482 m/s), which are due to soft $Te_7$ nets weakly embedded among the rigid $Re_6Se_8$ ($Re_6Te_8$) quasi-cubic clusters. The appearance of boson peak, i. e., hump of $C(T)/T^3$, verifies the existence of disordered phonon transports. Besides, the temperature dependence of $\kappa$ can be described by classic Debye-Callaway model. Notably, above 350 K, the $\kappa$ values of $Re_6Se_8Te_7$ and $Re_6Te_{15}$ are remarkably close to the upper limit derived from glassy-like diffusion model. This finding sets the superatomic compounds as a promising family for searching ultralow-$\kappa$ and energy management materials.

## 1. INTRODUCTION

In semiconductors and insulators, $\kappa$ is mainly determined by the lattice vibrations or phonons. The contribution of phonons can be described by the kinetic model, $\kappa = Cvl/3$, where $C$ is volumetric heat capacity, $v$ phonon speed, and $l$ phonon mean free path. To search intrinsic low-$\kappa$ single crystals, several ingredients of mismatching electron distribution, lattice vibration and chemical bonds are proposed so far in documented literatures. The first one is having strong anharmonic vibrations induced by the large deviations from the equilibrium positions of atoms. For example, the lone pairs induce the asymmetric distribution of electrons cloud, enhancing anharmonicity in SnSe,[1] PbTe,[2,3] $K_2Bi_8Se_{13}$,[4] and $Tl_3VSe_4$.[5] Secondly, the localized rattling-like vibrations due to atom vacancies, encaged atoms and disordered arrangements usually lead to low energy and soft mode phonons, which mixes the acoustic phonon and enhances the scattering. Such known compounds like $Ba_8Ga_{16}Ge_{30}$,[6,7] $CsAg_5Te_3$,[8] and $TlInTe_2$[9] show low $\kappa$ of 0.2-1.5 W $m^{-1}$ $K^{-1}$ at 300 K. Thirdly, the existence of anisotropic van der Waals interactions in layered compounds $Bi_4O_4SeCl_2$,[10] and $Cs_3Bi_2I_6Cl_3$[11] significantly reduces out-of-plane $\kappa$ to 0.11-0.20 W $m^{-1}$ $K^{-1}$, which is two or three times smaller than that of in-plane. In real materials, however, it is challenging to integrate two or three of the above-mentioned factors.

Superatomic compounds are composed of atomic clusters, where multiple atoms are connected together by ionic or covalent bonds, behaving as if there were individual atoms.[12] Due to their significantly enlarged size, the influence of these units on material properties is amplified, exhibiting rare physical properties.[13–17] From the perspective of crystal structure, the structural complexity is proportion to the numbers of atoms in a unit cell, which is higher than those of conventional atomic compounds, favoring low $\kappa$. In addition, if the clusters contain many heavy atoms, the low-frequency collective vibrations and much lower $\kappa$ can be highly anticipated. More interestingly, the clusters are always connected by weak chemical bonds, which are rather important to form a superatomic crystal. In case of inappropriate bonding strength, they will coalesce to form larger clusters or may even disintegrate into gaseous clusters.[18] In just a few relevant reports, the $\kappa$ in organic-inorganic superatomic $Co_6E_8(PEt_3)_6$ [E = S, Se, Te; Pet = poly(ethylene terephthalate)] is as low as 0.11 W m$^{-1}$ K$^{-1}$ at 300 K. The main reason is owing to its large and heavy [$Co_6E_8$] blocks are hierarchically linked by small and light organic ligands, leading to the $v$ as low as ~500 m/s.[19] The hard-soft mismatch is limited by size, charge transfer and bonding arrangements. It is hard to replace the ligands by other molecules because the total electron counts and structural symmetries obey a certain periodicity.[20,21] Another superatomic compound $Pt_3Bi_4Q_9$ [Q=S, Se] exhibits $\kappa$ of 0.61 W m$^{-1}$ K$^{-1}$ at 300 K, where the [$Pt_6Q_{12}$]$^{12-}$ clusters are connected by [$Bi_2Q_4$]$^{2-}$ units. The $\kappa$ is measured on polycrystalline hot-pressed pellet, which possibly compromises the intrinsic thermal property to some extent.[22] Given that superatomic materials contain numerous different atoms and types of chemical bonds, growing accessible bulk single crystals is extremely difficult. Therefore, studies on the intrinsic thermal transport properties of superatomic materials remain limited.

Previous works reported the $\kappa$ of ~1.3 W m$^{-1}$ K$^{-1}$ for $Re_6Te_{15}$ powders at 300 K, respectively.[23–26] Polycrystalline $Re_6Se_8Te_7$ was only reported some potential thermoelectric performance.[27,28] The absence of single crystals has hindered accurate measurements of intrinsic thermal properties. Herein, we firstly grow the bulk single crystal (2×2×2 mm$^3$) of superatomic $Re_6Se_8Te_7$ and $Re_6Te_{15}$ via high temperature $K_2Te$-flux method, enabling us to measure the intrinsic thermal properties. In both crystals, rigid quasi-cubic $Re_6Se_8$ and $Re_6Te_8$ clusters are interconnected through soft hinge-like $Te_7$ nets, which act as flexible and stretchable springs. They weaken inter-cluster interactions and induce strong anharmonicity and the low sound speed of 1420 m/s. The temperature dependent $\kappa$ follows the phonon-dominated picture in a single crystal below 300 K,[29] in which the $\kappa$ of $Re_6Se_8Te_7$ and $Re_6Te_{15}$ are respectively 0.32±0.02 W m$^{-1}$ K$^{-1}$ and 0.53±0.02 W m$^{-1}$ K$^{-1}$ at room temperature, and reduce to 0.26±0.02 W m$^{-1}$ K$^{-1}$ and 0.31±0.02 W m$^{-1}$ K$^{-1}$ at 390 K. These values are among the lowest ever observed in three-dimensional all-inorganic crystals. At $T>350$ K, the $\kappa$ of both crystals equal to the limit of Cahill-Pohl diffusion model,[30,31] highlighting the mismatch of vibration and bonding strength between rigid clusters and soft $Te_7$ net.[22,32,33]

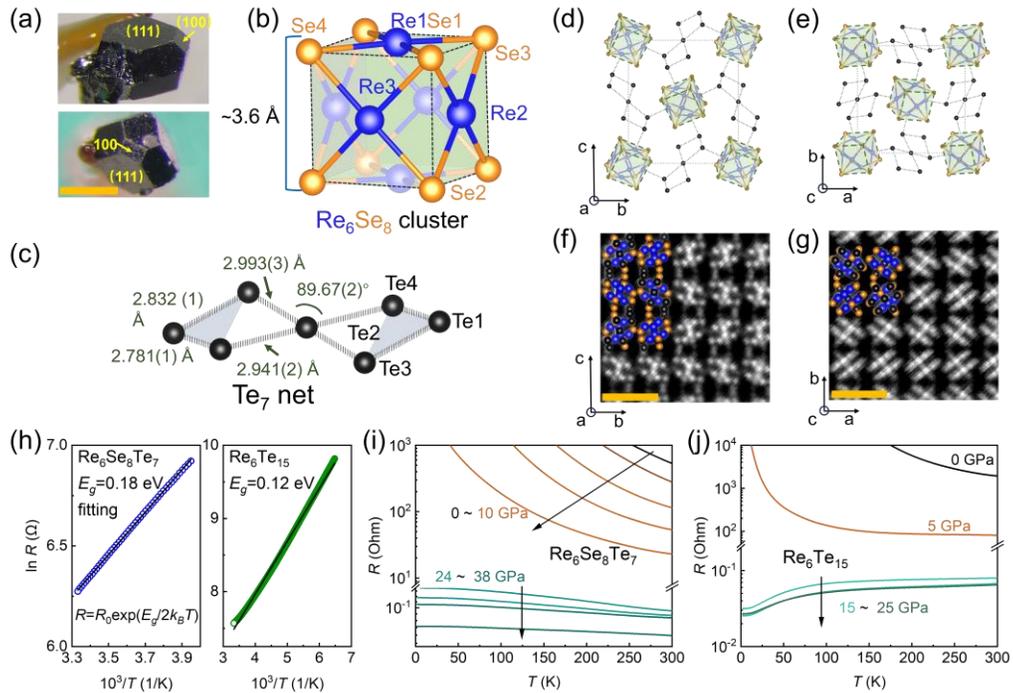

**Figure 1.** (a) Optical images of $Re_6Se_8Te_7$ (upper) and $Re_6Te_{15}$ (lower) with the hexagonal (111) and the quadrilateral (100) facets. The scale bar is 1 mm. (b) $Re_6Se_8$ cluster in geometry of a pseudo-cube and (c) the $Te_7$ net in which Te2, Te3 and Te4 are situated in a gray-colored plane. Crystal structure of $Re_6Se_8Te_7$ in the $bc$-plane (d) and $ab$-plane (e), respectively. (f) HAADF-STEM images of the $Re_6Se_8Te_7$ in (f) (100) and (g) (001) facet, together with the superimposed atoms (Re-blue, Se-orange, Te-black). The scale bar is 1 nm. Temperature-dependent electrical resistances (lnR vs. 10$^3$/T) of (a) $Re_6Se_8Te_7$ and (b) $Re_6Te_{15}$, respectively. The thermally activated band gaps ($E_g$) were determined by using the equation in (h). Temperature-dependent electrical resistances of $Re_6Se_8Te_7$ (i) and $Re_6Te_{15}$ (j) under pressure, respectively.

## II. RESULTS AND DISCUSSION

**A. Single crystal, electrical property and chemical bonding state.** The as-grown single crystals of $Re_6Se_8Te_7$ and $Re_6Te_{15}$ predominantly expose quadrilateral {100} and hexagonal {111} crystal facets, as shown in Figure 1a. The two samples are stable in air. Thermogravimetric analysis (TGA) shows that the two samples are thermally stable up to 750 K at $N_2$ atmosphere as shown in Figure S1. We determined the crystal structure through single-crystal x-ray diffractions at 300 K and 50 K. The refined bond lengths and bond angles and crystallographic parameters are listed in Table S1. Both compounds are isostructural showing the space group of *Pbca*. The lattice constants at 300 K of $Re_6Se_8Te_7$ are $a = 12.5579(3)$ Å, $b = 12.5488(4)$ Å, $c=14.0674(4)$ Å, and the counterparts of $Re_6Te_{15}$ are $a = 13.031(1)$ Å, $b = 12.949(2)$ Å, $c = 14.237(1)$ Å. For the former one, it consists of two components, pseudo-cubic $Re_6Se_8$ clusters (Figure 1b) and scissor-hinge-like $Te_7$ nets (Figure 1c). In the cluster, six blue Re atoms locate at the face centers of a pseudo-cubic, and eight orange Se atoms at the vertex. Strong Re-Se bonding is evidenced by their nearest-neighbor distance in the range of 2.524(9)-2.563(1) Å. In the $Te_7$ net, the bond length of Te1-Te3 is 2.781(1) Å, and the longest distance is 2.993(3) Å of Te2-Te3, suggesting the weak connections around Te2 atom. These bond lengths are slightly stronger than quasi-bonds in Te-Te dimers,[34,35] meanwhile, five Te atoms (Te2, 2×Te3, 2×Te4) form a planar configuration with a interchain angle of 89.67(3) °. This basal plane forms a 19.54(2) ° dihedral angle with the out-of-plane Te1-Te3-Te4 triad (Figure S2). In $Re_6Te_{15}$, all Se atoms at cubic vertex are replaced by Te, and the chemical bonds in $Te_7$ nets are almost identical to those of $Re_6Se_8Te_7$. The two compounds are both in a 3D structure in contrast to 0D $[Bi_2I_9]^{3-}$ in $Cs_3Bi_2I_9$,[36] due to the existence of Re-Te bonds between the $Re_6Se_8$ clusters and $Te_7$ nets. In addition, the detailed analysis of X-ray photoelectron spectra of $Re_6Se_8Te_7$ and $Re_6Te_{15}$ are shown in Figure S3. The valence can be assigned as $[Re_6Se_8]^{2+}$, $[Re_6Te_8]^{2+}$, and $[Te_7]^{2-}$ based on the peak position and previous references[37,38].

Figure 1d and e illustrate the crystal structure in *bc*- and *ab*-plane, respectively, where the clusters and nets are bridged by Re-Te bonds in the range of 2.6929(8)~2.7189(7) Å. This architecture resembles the NaCl-type alternating arrangement between $Re_6Se_8$ clusters and $Te_7$ nets. HAADF-STEM (high-angle annular dark-field scanning transmission electron microscopy) measurements validate the structure through matching atom and structural model. It reveals cluster periodicity with inter-cluster spacings of ~6 Å and ~9 Å along *a/b*- and *c*-axis, respectively. The hinge-like $Te_7$ behave like flexible springs among $Re_6Se_8$ clusters, which may induce strong anharmonic vibrations in three-dimensional space.

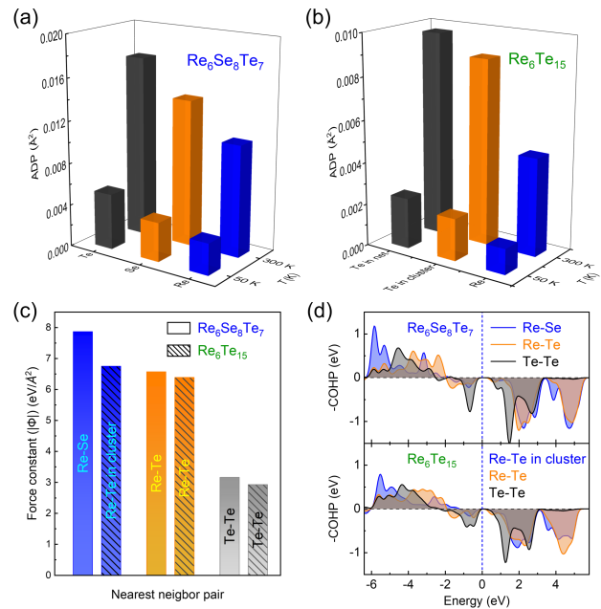

**Figure 2**. ADP value at 50 K and 300 K in (a) $Re_6Se_8Te_7$ and (b) $Re_6Te_{15}$. (c) IFC of the nearest neighbor Re-Se, Re-Te and Te-Te pairs in $Re_6Se_8Te_7$ and $Re_6Te_{15}$. (d) Average crystal orbital Hamilton population (-COHP) curves of the nearest atoms in $Re_6Se_8Te_7$ and $Re_6Te_{15}$.

In Figure 1h, electrical transport measurements exhibit semiconducting behavior at ambient. Their thermally-activated band gaps are 0.18 eV and 0.12 eV for $Re_6Se_8Te_7$ and $Re_6Te_{15}$, respectively, which are consistent with their optical band gaps of 0.17(1) eV and 0.10(1) eV through optical conductivity measurements as shown in Figure S4. With increasing external pressure, the *T*-dependent electrical resistance change into metallic like. The Seebeck coefficients of Re$_6$Se$_8$Te$_7$ are positive in 190-350 K, indicating a p-type semiconductor. The data above 300 K is close to the value in previous report, see Figure S5.[39]

To elucidate atomic coordination environments in Re$_6$Se$_8$Te$_7$ and Re$_6$Te$_{15}$, we refined the isotropic atomic displacement parameter (ADP) by $U_{iso} = h^2T/(4\pi mk_B\Theta_D^2)$, where $h$ is the Planck constant, $m$ the atomic mass, $k_B$ the Boltzmann constant, and $\Theta_D$ Debye temperature (Figures 2a and b). It can be seen that the Re atoms show minimal ADP value due to strong Re-Se bonds (Re-Te bonds for Re$_6$Te$_{15}$) and the Se atoms (Te atoms for Re$_6$Te$_{15}$) have slightly higher ADP value. Note that the Te atoms display the largest ADP at 50 K and 300 K, and the Te2 show higher ADP value (0.018 Å$^2$) for Re$_6$Se$_8$Te$_7$ as shown in Table S2, comparable to rattling magnitude of Ba in Ba$_8$In$_{16}$Ge$_{30}$, Ba$_8$Ga$_{16}$Ge$_{30}$, and Ba$_8$Ga$_{16}$Si$_{30}$,[8,9,40] which originates from the longest Te2-Te3 distance in Te$_7$ net.

The calculated harmonic interatomic force constants (IFCs) of the nearest-neighbor reveal that Te-Te interaction (3.16 eV/Å$^3$) is the weakest, being 52% and 60% smaller than the Re-Te (6.57 eV/Å$^3$) and Re-Se (7.87 eV/Å$^3$) bonds, respectively, as shown in Figure 2c. For Re$_6$Te$_{15}$, the IFCs of all atomic pairs are very close to those in Re$_6$Se$_8$Te$_7$. We performed a crystal orbital Hamilton population (COHP) analysis to quantitatively evaluate pairwise atomic interaction strengths. As shown in Figure 2d, all atomic pairs exhibit antibonding states below the Fermi level ($E_F$). In Re$_6$Se$_8$Te$_7$, comparing to other pairs, the Te-Te shows a high -COHP value of 0.76 eV below the $E_F$. As for Re$_6$Te$_{15}$, the -COHP value of Te-Te lightly decreases below the $E_F$. Based on integrated COHP value (-iCOHP) near $E_F$ shown in Figure S6, we can see that all the Te-Te bonds are located in weak regime. Furthermore, these antibonding valence states could lower the frequency acoustic phonon modes and weaken bond strength, thus reducing the $\kappa$.[41]

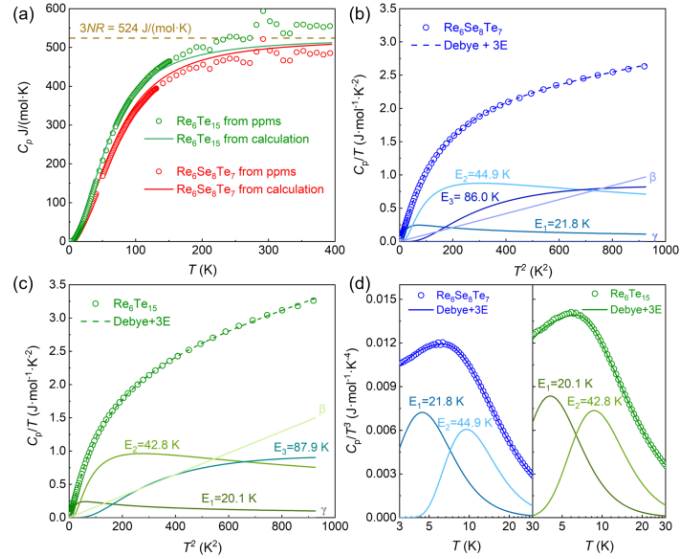

**Figure 3**. (a) Temperature-dependent heat capacity ($C_p$) of Re$_6$Se$_8$Te$_7$ and Re$_6$Te$_{15}$ in the range of 2-400 K. Circles and solid lines are the measured and calculated data, respectively. Heat capacity $C_p$ of (b) Re$_6$Se$_8$Te$_7$ and (c) Re$_6$Te$_{15}$ measured in the range of 2-30 K presented as $T^2$ dependent $C_p/T$ curves, respectively. (d) Low-temperature $C_p/T_3$ vs $T$. The positions of the peak are consistent with corresponding Einstein modes.

**B. Heat capacity and anharmonic vibration.** Heat capacity measurements serve as an effective probe for anharmonic vibrations. The $C_p$ above 250 K saturate to the classical value of 524 J/mol K (see Figure 3a). As shown in Figures 3b and c, the $C_p/T$ -$T^2$ plots are presented for Re$_6$Se$_8$Te$_7$ and Re$_6$Te$_{15}$, respectively, with well-fitted results by the Debye-Einstein model

$$\frac{C_p}{T} = \delta + \beta T^2 + \sum_i A_i(E_i)^2(T^2)^{\left(-\frac{3}{2}\right)}\frac{e^{E_i/T}}{(e^{E_i/T}-1)^2} \quad (1)$$

In equation 1, $\delta$ is the linear term that originates from electrons or glassy structures, and the $\beta=\eta(12\pi^4R/5)(\Theta_D)^{-3}$ is associated with the lattice contribution of those harmonic vibrations in Debye modes ($\eta=1-\sum_iA_i/3NR$). Here, $R$, $\Theta_D$ and $N$ are the universal gas constant, Debye temperature, and the number of atoms per formula unit, respectively. The last term represents contribution of Einstein oscillators, i.e., those soft and anharmonic vibration modes,[42] where $A_i$ and $E_i$ represents the pre-factor and Einstein temperature of the $i$th Einstein mode, respectively. The fitted parameters are listed in Table S3. As depicted in Figures 3b and c, the curves are well fitted by adding three Einstein oscillators of $E_1$=21.8 K, $E_2$=44.9 K and $E_3$=86.0 K for $Re_6Se_8Te_7$ and $E_1$=20.1 K, $E_2$=42.8 K, and $E_3$=42.8 K for $Re_6Te_{15}$. The $\Theta_{DS}$ of $Re_6Se_8Te_7$ and $Re_6Te_{15}$ are 154.9 K and 144.4 K, respectively. We plot the $C_p/T^3$-$T$ curves in Figure 3d. The data can be reproduced by the Debye-Einstein model. More importantly, the humps at ~7 K are signatures of boson peaks, which stem from weakened vibration of $Te_7$.[43–46] It is known that boson peaks are usually observed in disordered glassy or amorphous materials. The appearance is a rare case, which indicates the competition between phonon propagation and diffusive-like damping associated with the extra anharmonic vibrations beyond Debye modes is strong.[47]

The phonon density of states (PDOS) in Figure 4a reveals that all vibrations are below 245 cm$^{-1}$, a consequence of heavy elements and weak inter-cluster coupling. Raman data >20 cm$^{-1}$ at several temperatures agree well with the phonon DOS, as shown in Figure S7. The frequency of 140-245 cm$^{-1}$ come from Re and Se (Te) atoms in clusters. The corresponding frequency was calculated from the Einstein temperature using the formula $\hbar\omega = k_BT$, where $\hbar$ is the reduced Planck constant and $k_B$ is the Boltzmann constant. In Figure 4b and 4c, two Einstein modes ($E_1$=15.13 cm$^{-1}$, $E_2$=31.10 cm$^{-1}$ for $Re_6Se_8Te_7$ and $E_1$=13.91 cm$^{-1}$, $E_2$=29.69 cm$^{-1}$ for $Re_6Te_{15}$) correspond to collective vibrations among the clusters with different orientations. They possibly relate to the origin of boson peaks. Notably, the $E_2$ mode can be observed as a characteristic peak in the Raman spectra of both $Re_6Se_8Te_7$ and $Re_6Te_{15}$, confirming the reliability of the calculations. Furthermore, the Te atoms in the $Te_7$ net contribute to the modes of 45–70 cm$^{-1}$, covering the Einstein modes $E_3$=59.6 cm$^{-1}$ and $E_3$=60.98 cm$^{-1}$ of $Re_6Se_8Te_7$ and $Re_6Te_{15}$, respectively. Detailed analyses show that these frequencies correspond to the collective breathing-like vibrations of the $Te_7$ net, as illustrated in Figure 4d.

Furthermore, the anharmonicity is quantitatively described using the Grüneisen parameter $\gamma_i$=-$\partial\ln\omega_i/\partial\ln V$, where $\omega_i$ and $V$ represent phonon frequency, and unit cell volume, respectively. Since the $\gamma_i$ is temperature-independent at high temperatures, and the average Grüneisen parameter is estimated by the equation $\gamma=B_0\alpha V/C_V$, where $B_0$, $\alpha$, $V$, and $C_V$ correspond to the bulk modulus, the volumetric thermal expansion coefficient, unit cell volume, and heat capacity at constant volume, respectively. We measured pressure and temperature dependent *in-situ* XRD to determine the $B_0$ and $\alpha$ of 300 K, see Figure 5a and 5b. Detailed XRD patterns are as shown in Figure S8 and S9. The fitted $B_0$ is 47.6 GPa using Birch-Murnaghan (BM) equation of state. The $\alpha$=4.8×10$^{-5}$ K$^{-1}$ derived from the slope of the fitted curve, following $\alpha$=($V$-$V_{300K}$)/($\Delta T\times V_{300K}$). The $C_V$ was calculated using the relation,[48] $C_V$=$C_P$-9$VTB\alpha^2$=390 J mol$^{-1}$ K$^{-1}$. Therefore, the derived $\gamma$ is 1.93 for $Re_6Se_8Te_7$, which is comparable to the value in typical IV-VI materials with strong anharmonic scattering,[2,49,50] superatomic $Co_6Se_8(PEt_3)_6$ ($\gamma$ is 1.78) and $Pt_3Bi_4Se_9$ ($\gamma$ is 1.97).[19,22]

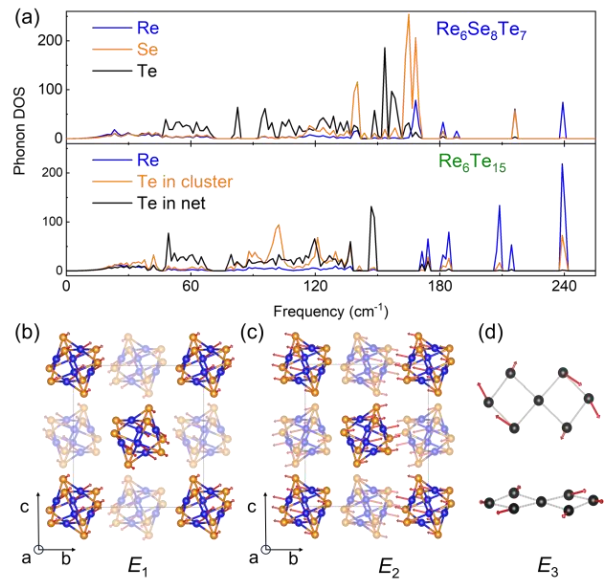

**Figure 4.** (a) Calculated projected phonon density of states for atoms in $Re_6Se_8Te_7$ and $Re_6Te_{15}$, respectively. Visualizations of Einstein oscillator modes of (b) $E_1$, (c) $E_2$ and (d) $E_3$ at the Γ point. The black box represents the primitive unit cell. Blue, orange, and black spheres represent Re, Se (Te in cluster for $Re_6Te_{15}$), and Te atoms, respectively.

We calculated the vibration frequency dependence of the $\gamma$, as shown in Figure 5c. The $\gamma$ of 15–30 cm$^{-1}$, aligning with the frequencies of the two Einstein modes that primarily contribute to the boson peak, range from 2.0-2.5. This reinforces the conclusion that the collective oscillations of the superatomic clusters play a crucial role in the anharmonicity due to the loosened inter-cluster connection via Te$_7$ nets. The anharmonicity originated from collective oscillations should be more significant in Re$_6$Se$_8$Te$_7$ because Re$_6$Se$_8$ clusters behave more like rigid and integral units than Re$_6$Te$_8$ clusters. Thus, Re$_6$Se$_8$Te$_7$ exhibits stronger anharmonic effects, leading to higher $\gamma$ than Re$_6$Te$_{15}$ as shown in Figure 5c. Furthermore, from the low-frequency phonon spectrum, it is observed that at ~15 cm$^{-1}$, the optical modes ($E_I$=15.13 cm$^{-1}$ for Re$_6$Se$_8$Te$_7$ and $E_I$=13.91 cm$^{-1}$ for Re$_6$Te$_{15}$) couple with acoustic modes, contributing to the boson peaks as highlighted by the green circles in Figure 5d. The coupling usually enhance anharmonicity and impede heat propagation, similar to the effect in CsAg$_5$Sb$_3$, TlInTe$_2$, and CuBiSCl$_2$.[8,9,22,50,51]

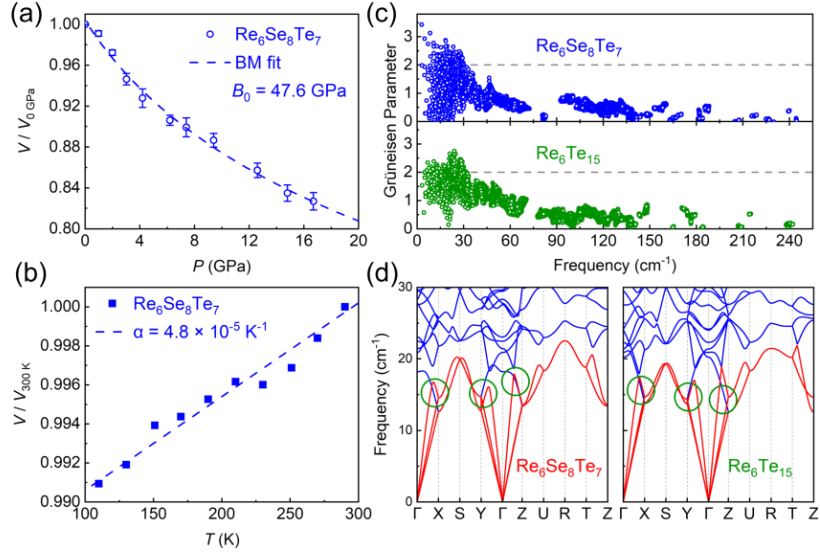

**Figure 5**. (a) Pressure dependent normalized crystal cell volume ($V/V_{0\,GPa}$) of Re$_6$Se$_8$Te$_7$. (b) Temperature dependent normalized $V/V_{300K}$ of Re$_6$Se$_8$Te$_7$. (c) Theoretical Grüneisen parameters as distributed over whole frequency. (d) Phonon dispersion in range of below 30 cm$^{-1}$. Green circles highlight the crossing-over between acoustic and optical branches.

**C. Thermal conductivity and sound velocity.** Figure 6a displays the measured $\kappa$ as a function of temperature, exhibiting ultralow $\kappa$ of 0.32±0.02 W m$^{-1}$ K$^{-1}$ and 0.53±0.02 W m$^{-1}$ K$^{-1}$ in Re$_6$Se$_8$Te$_7$ and Re$_6$Te$_{15}$ at room temperature, respectively. The representative frequency-domain thermoreflectance (FDTR) data and best-fit curves are shown in Figure S10.[52,53] The electronic thermal conductivity $\kappa_e$ of two samples are shown in Figure S11. Since the electronic contribution accounts for less than 0.01% of the total $\kappa$, the $\kappa_e$ can be neglected. The $\kappa$ of Re$_6$Se$_8$Te$_7$ is lower than that of Re$_6$Te$_{15}$ because Re$_6$Se$_8$Te$_7$ has stronger anharmonic effects, consisting with the trend in Se substituted Re$_6$Te$_{15-x}$Se$_x$ powders.[26,25,24] However, comparing with previous results $\kappa \approx 1.3$ W m$^{-1}$ K$^{-1}$ and $\kappa \approx 0.5$ W m$^{-1}$ K$^{-1}$ for Re$_6$Te$_{15}$ and Re$_6$Se$_6$Te$_6$ polycrystalline samples at 300 K, the crystal samples even possess lower $\kappa$, indicating the primary contribution to $\kappa$ of intrinsic anharmonicit. Both $\kappa$ follow the inverse temperature dependence characteristic of Umklapp scattering-dominated phonon transport within the Peierls-Boltzmann framework. To gain further insights into the ultralow $\kappa$, we fit the $\kappa$-$T$ curve using a modified Debye-Callaway model[10,33,54] as

$$\kappa_L(T) = \frac{\hbar^2}{2\pi^2 v k_B} \int_0^{\omega_{max}} \frac{\omega^4}{T^2} \cdot \frac{1}{\tau_{ph}^{-1}} \cdot \frac{e^{\frac{\hbar\omega}{k_B T}}}{\left(e^{\frac{\hbar\omega}{k_B T}} - 1\right)^2} d\omega \quad (2)$$

Here, $\omega_{max}$ is the Debye cut-off frequency $\omega_{max}=\Theta_D/k_BT$, where $\Theta_D$ is the Debye temperature, $k_B$ the Boltzmann constant, $\hbar$ the reduced Planck constant, $\omega$ the phonon angular frequency, $v$ the average velocity of sound, and $\tau_{ph}$ the phonon relaxation time, respectively. According to Matthiessen's rule,[55] the scattering rate $\tau_{ph}^{-1}$ is a sum of various scattering process,

$$\tau_{ph}^{-1} = \tau_B^{-1} + \tau_D^{-1} + \tau_U^{-1} = \frac{v}{L} + A\omega^4 + \frac{\hbar\gamma^2}{Mv^2\Theta_D}\omega^2 T e^{-\frac{\Theta_D}{3T}} \quad (3)$$

$\tau_B$, $\tau_D$, and $\tau_U$ are the boundary, point-defect, and Umklapp scatterings, respectively. $M$ is average atomic mass and $\gamma$ Grüneisen parameter. Considering the high quality of our single crystals, we ignore the influence of boundaries and point defects, and only use the Umklapp term. The limited-memory Broyden Fletcher Goldfarb Shanno (L-BFGS) optimization algorithm was then employed to fine-tune $v$ and $\gamma$ during the best fitting process. The fitted $v$ is 1400 m/s for both compounds, and $\gamma$ is 4 ($Re_6Se_8Te_7$) and 3 ($Re_6Te_{15}$), respectively. Besides, the phonon lifetimes and mean free paths (MFPs) as functions of frequency were computed for both compounds at 300 K (Figure S12). For $Re_6Se_8Te_7$, at $T>300$ K, the $\kappa$ is lightly higher than the Debye-Callaway model. This may be attributed to the two-channel propagation since the MFPs of some phonon fall below the Ioffe-Regel limit at superatomic scale (see Figure S12).[5,56] We fit the full width at half maximum (FWHM) of the $E_2$ mode in $Re_6Se_8Te_7$, and analyzed its temperature dependence with a semi-quantitative simple model (see Figure S13)[57,58]. The fitting result implies that the possible four-phonon scattering processes induce deviations from the Debye model at high temperatures. At 300 K, the extracted FWHM corresponds to a 4.4 ps phonon lifetime, matching the result in Figure S8. Besides, We calculated the minimum thermal conductivity based on the Cahill-Pohl model (see Equation S1).[30,31,56] Notably, above 350 K, both $\kappa$ closely approach this saturated value.

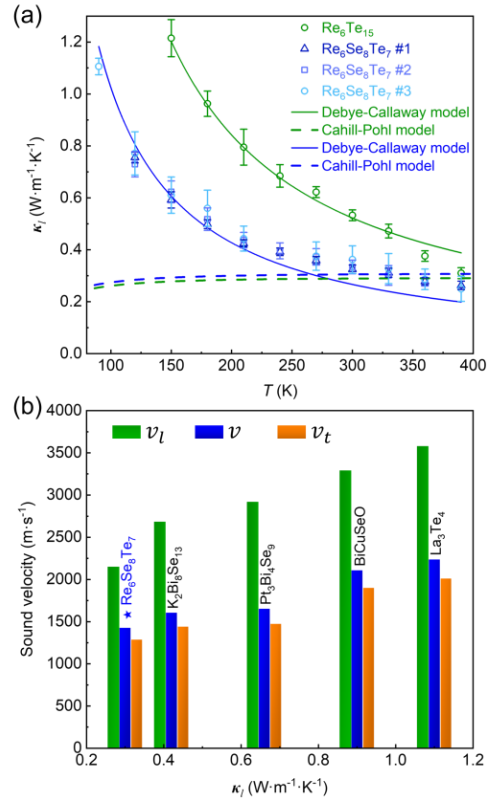

**Figure 6**. (a) Temperature-dependent $\kappa$ of $Re_6Se_8Te_7$ and $Re_6Te_{15}$. The data can be fitted by the Debye-Callaway model (solid lines). The dashed lines are calculated curves based on the Cahill-Pohl diffusion model. (b) Bar chart of sound velocity ($v$), and $\kappa$ of $Re_6Se_8Te_7$ and several known materials governed by the Umklapp process. The green, blue and orange bars indicate longitudinal ($v_l$), average ($v$), and transverse ($v_t$) sound velocity, respectively.

We also measured the sound velocity of $Re_6Se_8Te_7$ and $Re_6Te_{15}$ by pulse-echo method, and listed them along with theoretical value in Table 1. The transverse sound velocities ($v_t$) of $Re_6Se_8Te_7$ and $Re_6Te_{15}$ are 1290 m/s and 1345 m/s, respectively, while their longitudinal sound velocities ($v_l$) are 2145 m/s and 2110 m/s. The corresponding average sound velocities ($v$) are 1425 m/s and 1478 m/s, which are remarkably low comparing with BiCuSeO (2107 m/s), $Pt_3Bi_4Se_9$ (1651 m/s) and $K_2Bi_8Se_{13}$ (1605 m/s). $Re_6Te_{15}$ has the larger average atomic mass and weaker intra-cluster bonding strength, but it has almost identical $v$.[11,48] According to the equation $v \propto (S/m)^{-0.5}$, where $S$ and $m$ represent elastic constant and average atomic mass, respectively. It implies that the $v$ here is mainly be determined by the $S$. Figure 6b shows the $\kappa$ and $v$ of $Re_6Se_8Te_7$ with low-thermal-conductivity materials governed by the Umklapp process.[22,48,59] The $Re_6Se_8Te_7$ has the lowest $v$, especially for longitudinal sound speed $v_l$. Herein, this low $v$ is ascribed to the scissor-hinge-like Te$_7$ nets, which weaken the opposing motion between superatomic clusters and lead to small $v_l$.

**Table 1.** Experimental sound velocity ($v_t$, $v_l$ and $v$) of Re$_6$Se$_8$Te$_7$ and Re$_6$Te$_{15}$ at 300 K measured by using pulse-echo method and theoretically calculated value (#Cal.).

|  | Re$_6$Se$_8$Te$_7$ #sample 1 | Re$_6$Se$_8$Te$_7$ #sample 2 | Re$_6$Se$_8$Te$_7$ #Cal. | Re$_6$Te$_{15}$ #sample 1 | Re$_6$Te$_{15}$ #sample 2 | Re$_6$Te$_{15}$ #Cal. |
|---|---|---|---|---|---|---|
| $v_t$ (m s$^{-1}$) | 1294.38 | 1282.85 | 1210 | 1342.66 | 1347.37 | 1200 |
| $v_l$ (m s$^{-1}$) | 2133.33 | 2165.41 | 2150 | 2098.36 | 2121.55 | 2200 |
| $v$ (m s$^{-1}$) | 1430.32 | 1420.87 | 1347 | 1475.18 | 1481.62 | 1340 |

There are only a few reports regarding to the ultralow $\kappa$ in superatomic compounds. In a newly-reported Pt$_3$Bi$_4$Q$_9$ (Q=S, Se), the $\kappa$ of polycrystalline pellet is nearly twice that of Re$_6$Se$_8$Te$_7$ at 300 K.[22] In C$_{60}$, the insert of K atoms among C$_{60}$ weakens the connections, which in turn reduces the $\kappa$ from 0.7 W m$^{-1}$ K$^{-1}$ of C$_{60}$ to 0.32 W m$^{-1}$ K$^{-1}$ of K$_3$C$_{60}$, and further to 0.21 W m$^{-1}$ K$^{-1}$ of K$_6$C$_{60}$.[60–62] Furthermore, the $\kappa$ can be reduced as low as 0.07 W m$^{-1}$ K$^{-1}$ by embedding organic molecule among C$_{60}$, a tenfold decrement compared with pristine C$_{60}$.[63,64] Thus, we highlight the significance of connection way between superatomic units for ultralow $\kappa$. In a straightforward comparison, in Re$_6$Se$_8$Cl$_2$ and Re$_6$Se$_8$I$_2$ where the [Re$_6$Se$_8$] clusters are linked by single Cl or I atom, the $\kappa$ is of 0.4 or 1.2 W m$^{-1}$ K$^{-1}$ at 300 K, respectively.[65,66] Synergic effect of large mass of [Re$_6$Se$_8$] cluster and asymmetric displacement of Cl or I atoms enhance the phonon scattering, thereby exhibiting low heat transport ability.

In Re$_6$Se$_8$Te$_7$ and Re$_6$Te$_{15}$, the Te$_7$ nets increase inter-superatomic distances to 6 Å and 9 Å. This structural feature can significantly enhance collective vibrations among superatomic clusters and weaken the stiff of lattice through its loose connections. The introduction of mismatched vibrational and bonding strength not only induce anharmonic and disorder effects, but also significantly reduces the propagation velocity of heat-carrying phonons, particularly longitudinal phonons along the contraction direction of Te$_7$ nets. At T> 350 K, the vibrations of Te atoms in Te$_7$ net become more vigorous. Given that the $l$ is smaller than the Ioffe-Regel limitation, the $\kappa$-$T$ should be described by the Cahill-Pohl model as previously-reported literature[56], and the $\kappa$ gradually approaches to the saturated value of this glass model.

### III. CONCLUSIONS

In summary, the importance of soft Te$_7$ nets to induce ultralow $\kappa$ in superatomic Re$_6$Se$_8$Te$_7$ and Re$_6$Te$_{15}$ is demonstrated by comprehensive experimental and theoretical characterizations. We would like to emphasize that the creating mismatch in lattice vibrational and chemical bonding may be applicable to many superatomic and other complex structure materials. To achieve intrinsically low $\kappa$, as proven herein, tuning the atomic numbers and chemical environments within local structure units represents a practical approach.

### IV. EXPERIMENTAL SECTION

**A. Single crystal growth.** Powders of Re$_6$Se$_8$Te$_7$ and Re$_6$Te$_{15}$ were grown using a solid-state reaction referring to previous reports.[27,28] Tellurium powder (99.99 %), selenium powder (99.99 %), and rhenium powder (99.9 %) were mixed in a stoichiometric ratio and loaded into an alumina crucible. The mixture was sealed in an evacuated quartz ampoule heated to 1000 °C over 48 h and maintained for 60 h before quenching in air. Then, single crystals of Re$_6$Se$_8$Te$_7$ and Re$_6$Te$_{15}$ were synthesized as follows: powder of Re$_6$Se$_8$Te$_7$ or Re$_6$Te$_{15}$ was used as the precursor and combined with a potassium lump (used as flux) and tellurium powder (as flux) in a molar ratio of 1:4:40 and loaded into an alumina crucible. The setup was arranged with the potassium lump at the bottom, the pressed Re$_6$Se$_8$Te$_7$ or Re$_6$Te$_{15}$ powder pellet in the center, and the tellurium powder on top. The mixture was sealed in an evacuated quartz ampoule heated to 400°C over 12 h and kept for 24 h. It was then slowly ramped up to 1000°C in 24 h, held at 1000°C for 36 h, and subsequently cooled down to 600°C over 72 h before centrifugation.

**B. Structure determinations.** Temperature-dependent *in-situ* Powder X-ray Diffraction Patterns (PXRD) were collected on a Rigaku Smart Lab diffractometer with Cu-$K_\alpha$ radiation ($\lambda$=1.5406 Å, 40 kV, 30 mA) and a germanium monochromator in reflection mode. The cooling process was carried out using a liquid nitrogen apparatus. *In-situ* high-pressure PXRD were performed at a Rigaku XtaLAB Synergy

R diffractometer using a diamond anvil cell (DAC) with a facet diameter of 300 μm, in which the source was multilayer mirror monochromatized Mo-$K_α$ ($λ$=0.71073 Å) radiation. Mineral oil served as the pressure-transmitting medium, and the pressure was calibrated by the standard ruby fluorescence method. Structure refinements of PXRD were performed using the Le Bail method and Rietveld analysis with FullProf software.[67] Visualization of crystal structures was illustrated using VSETA software.[68,69]

Single crystal diffraction data were collected on a Bruker D8 Venture diffractometer equipped with a Photon-II detector. A graphite monochromatized Mo-$K_α$ radiation ($λ$=0.71073 Å) was used for the diffraction experiment. Data collection, cell refinement, and data reduction were carried out in the Bruker APEX5 program. The data were solved using the Olex2 software package.[70]

Transmission electron microscopy (TEM), ED, high-angle annular dark-field scanning transmission electron microscopy (HAADF-STEM), and energy dispersive x-ray spectroscopy (EDXS) elemental mapping experiments were carried out on double (probe and image) aberration corrected cold FEG JEM ARM200F microscope operated at 200 kV and equipped with a CENTURIO EDX detector, an ORIUS Gatan camera and Quantum GIF. The TEM specimen was prepared by mechanical grinding the samples in ethanol and depositing the obtained suspension on Ni carbon holey grid.

**C. Electrical transport.** *In-situ* high-pressure electrical transport of the single crystals was measured in a physical property measurement system (PPMS, Quantum Design, DynaCool). Transport measurements were carried out using a beryllium copper alloy DAC with a facet diameter of 300 μm, and pressure was monitored with the ruby fluorescence method. We use KBr as the pressure-transmitting medium.

**D. Heat capacity and thermal conductivity measurements.** The heat capacity was collected on Quantum Design Physical Property Measurement System using single crystal, with a temperature range of 2−400 K. The thermal conductivity was measured by frequency-domain thermoreflectance (FDTR). Our platform is based on two continuous-wave (CW) lasers. The pump laser (488 nm) is sinusoidally modulated from 100 kHz to 20 MHz to heat the sample surface and another laser (532 nm) probe the surface temperature change. The effective laser spot size (root-mean-square average of the pump and probe beam radii) in this experiment is 2.9 μm with a 10× objective. The signal of the reflected probe beam is detected by a balanced detector connected with the radio-frequency lock-in amplifier. To extract the sample thermal conductivity, the phase signal was fitted to a Fourier heat conduction model (Figure S6).

**E. Raman spectrum.** The spectra were obtained by using a WITec alpha 300R instrument in a quasi-backscattering geometry with a solid-state laser excitation line at $λ$ = 532 nm. The laser beam, with a power of 0.5 mW and a duration of 120 s, was focused to a 3 μm diameter spot on the surface of the samples.

**F. Seebeck Coefficient.** The Seebeck coefficient were measured using a Thermal Transport Measurement System (TTMS, Multifields Tech.) with sintered $Re_6Se_8Te_7$ powder. The steady-state method was employed for measuring the total thermal conductivity and Seebeck coefficient.

**G. Sound Velocity.** The longitudinal and transverse sound velocities at 300 K were measured using the pulse-echo method on a UMS-100 instrument.

**H. Theoretical Calculations.** First-principles calculations were carried out with the density functional theory (DFT) implemented in the Vienna ab initio simulation package (VASP)[71]. The generalized gradient approximation (GGA) in the form of Perdew-Burke-Ernzerhof (PBE)[72] was adopted for the exchange-correlation potentials. We used the projector augmented-wave (PAW)[73] pseudopotentials with a plane wave energy of 400 eV; $5p^65d^66s^1$ for Re, $5s^25p^4$ for Te, and $4s^24p^4$ for Se were treated as valence electrons, respectively. Only Γ point was used in the self-consistent calculation considering the large number of atoms in the unit cell. The self-consistent field procedure was considered converged when the energy difference between two consecutive cycles was lower than $10^{-6}$ eV. Atomic positions and lattice parameters were fully relaxed till all the forces on the ions were less than $10^{-3}$ eV/Å. The Crystal Orbital Hamilton Population (COHP) analysis was carried out with LOBSTER package,[74] and a Γ-centered Monkhorst-Pack Brillouin zone sampling 0.02 ×2π Å$^{-1}$ was adopted.[9] Phonon spectra were calculated on 1×1×1 supercells for both $Re_7Te_{15}$ and $Re_6Se_8Te_7$ using the finite displacement method implemented in the PHONOPY code.[75] The grüneisen parameter γ was calculated by

$$\gamma_i = -\frac{V}{\omega_i}\frac{\partial \omega_i}{\partial V},$$

where the volume was isotropically expanded by ±1% in this work. The interatomic force constant was calculated by Phonopy.[75]

**SUPPLEMENTARY MATERIALS**

See the supplementary materials for Equation S1. Figures show atomic coordinations and bond lengths, detailed XRD patterns, Raman spectra, and representative FDTR fitting results. Tables present single-crystal refinement results, ADP values, and heat capacity fitting results.

## ACKNOWLEDGEMENTS

We thank Xu Chen, Xiaowei Wu, and Qi Li for fruitful discussions. This work is supported by the National Natural Science Foundation of China (52250308), the National Key Research and Development Program of China (2023YFA1406301, 2024YFA1207900), the National Natural Science Foundation of China (52525205, 52202342, 12374012) and Shandong Provincial Natural Science Foundation (ZR2023JQ001). This work was supported by the Synergetic Extreme Condition User Facility (SECUF, https://cstr.cn/31123.02.SECUF)-High-pressure Synergetic Measurement Station, the Sample Pre-selection and Characterization Station, and the Infrared Unit in THz and Infrared Experimental Station.

## AUTHOR DECLARATIONS

### Conflict of Interest

The authors declare no competing financial interest.

### Author Contributions

‡These authors contributed equally. X. L., J.G., B.S., and X.C. conceptualized the research idea and supervised the project. M.Y. synthesized the samples. M.Y. and Y.W. conducted the physical property characterizations and data processing. J.D. performed the theoretical calculations. Y.W. measured the thermal conductivity. Q.Z. measured the STEM images. X. L., J. G., M.Y., Y.W., N. L., B.S., and X. C. analyzed the data and wrote the paper based on the discussion with all authors.